# SIMULATION OF DISTRIBUTED MANUFACTURING ENTERPRISES: A NEW APPROACH

Sameh M. Saad
Terrence Perera
Ruwan Wickramarachchi

School of Engineering
Sheffield Hallam University
Sheffield, S1 1WB, U.K.

## ABSTRACT

The globalization of markets and world-wide competition forces manufacturing enterprises to enter into alliances leading to the creation of distributed manufacturing enterprises. Before forming a partnership it is essential to evaluate viability of proposed enterprise as well as how a company's operations are affected by the proposed virtual enterprise. Distributed simulation provides an attractive tool to make decisions on such situations. However, due to its complexity and high cost distributed simulation failed to gain a wide acceptance from industrial users. This paper presents a new approach for distributed manufacturing simulation (DMS) including a formal methodology for DMS and, implementation approach using current commercial simulation software, employing widely available and cost effective technologies. The main objective of this work is to promote the use of distributed simulation particularly in distributed manufacturing by making it fast to develop and less complicated for implementation.

## 1 INTRODUCTION

Confronted with growing competition, the evolution of new markets and increasingly complex global and political scenarios, today's manufacturing organizations are forced to rethink about how they are organized and operated. Not only to gain a competitive advantage over their competitors but often merely to survive, companies are now looking for innovative ways to respond to market changes, produce better quality products in more cost effective manner, manage product life cycles effectively etc. As a result, enterprises are moving towards more open architectures for integrating their activities with those of their suppliers, customers and partners within wide supply chain networks (Shen and Norrie 1998). In manufacturing, companies may form strategic partnerships for outsourcing some of their operational activities, share resources or joint development of products and services etc., leading to formation of virtual manufacturing enterprises which operate in distributed manufacturing environment. To facilitate the creation of virtual manufacturing enterprises, potential partners must be quickly able to evaluate whether it will be profitable for them to participate in the proposed enterprise. Simulation provides a capability to conduct experiments rapidly to predict and evaluate the results of manufacturing decisions (McLean and Leong 2001).

Simulation is not a strange tool for decision making in manufacturing. Law and McComas (1998) pointed out that manufacturing is one of the largest application areas of simulation, with the first uses dating back to at least early 1960s. However, traditional sequential simulation alone may not sufficient to simulate these highly complex Distributed Manufacturing Enterprises (DMEs). In such situations, distributed simulation provides a promising alternative to construct cross enterprise simulations. Each partner can use a simulation of its operation to make sure that it has the capability to perform its individual function in the DME. Later these simulation models can be integrated into a distributed simulation to simulate the whole enterprise, and evaluate the feasibility and profitability of the proposed partnership. The use of distributed simulation allows each partner to hide any proprietary information in the implementation of the individual simulation, simulate multiple manufacturing systems at different degrees of abstraction levels, link simulation models built using different simulation software, to take advantage of additional computing power, simultaneous access to executing simulation models for users in different locations, reuse of existing simulation models with little modifications etc. (Gan et al. 2000; McLean and Riddick 2000; Taylor et al. 2001; Venkateswaran et al. 2001). However, Peng and Chen (1996) noted that as a technique, parallel and distributed simulation is not successful in manufacturing. Most of the simulations for DMEs implemented so far are purpose build simulators created using programming languages such as C++ or Java, and with high end computers. Furthermore distributed simulation itself involves long development time, cost, steep learning curves and complex to manage, and is also criticized for lack of penetration into industrial applications. Implementation of a distributed simulation requires not only expertise on distributed simulation, but also expertise in programming languages too.

This paper presents a new distributed simulation approach using commercial simulation software to implement DMS. The reason to propose commercial simulation software is that most of the companies already use commercial simulation software in manufacturing applications. Our objective is to use widely available tools and technologies that are familiar to existing users or easy to learn, in order to develop a distributed simulation system in cost (also time) effective manner. The proposed approach includes a formal methodology to develop DMS and a synchronization mechanism to implement distributed manufacturing simulation.

## 2 BACKGROUND

### 2.1 Distributed Manufacturing

Throughout the century, the world of manufacturing has changed from a mainly in-house effort to a distributed style of manufacturing. As the term distributed manufacturing implies, distributed manufacturing enterprises which also known as virtual manufacturing enterprises operate in geographically distributed environment and connected together with modern communication technologies. Virtual manufacturing enterprises are ephemeral organizations in which several companies collaborate to produce a single product or product line (Venkateswaran et al. 2001). Participating in this type of collaboration allow partner organizations to use their knowledge, resources and in particular manufacturing expertise to take advantage of new business opportunities and/or gain a competitive advantage that are on a larger scale than an individual partner could handle alone. Generally these type of enterprises are established without making a long term commitment to other partners and individual partners may also carry out their own manufacturing activities independent of activities relating to the DME.

Due to their nature and also to the environment they operate, DMEs are highly complex and heterogeneous. The traditional manufacturing control systems have low capacity to adapt and to react to the complex and dynamic nature of DMEs. Therefore, attempts have made to develop distributed manufacturing architectures that can deal with complex and dynamic systems. New control and organizational architectures such as Agile, Fractal, Bionic, Random, responsive manufacturing, and Holonic manufacturing architectures have been introduced over the last few years (see Kadar et al. 1998; Leitao and Resviti 2000; Leitao and Resviti 2001; Saad 2003 for more details).

### 2.2 Distributed Simulation

Distributed simulation combines distributed computing technologies with traditional sequential simulation techniques. Although with drawbacks such as Low communication speeds and shortage of network bandwidth, its popularity increased in recent years due to availability of powerful but low cost desktop workstations and improvements in networking technologies. In distributed simulation, the simulated system is partitioned or decomposed into a set of subsystems that are simulated in interconnected workstations. Bargrodia (1996) viewed distributed simulation as a collection of sequential simulation models, which communicate each other with timestamped messages. A Synchronized simulation system makes sure that each individual simulation model processes arriving messages in their timestamped order and not in real time arriving order. This requirement is referred to as local causality constraint (Fujimoto 1999). To satisfy the local causality constraint, number of synchronization protocols have been proposed in the literature. These protocols can be broadly classified as conservative or optimistic protocols (Fujimoto 1990). Conservative approaches strictly impose the local causality constraint and guarantee that each model will only process events in non-decreasing timestamp order. In contrast, optimistic approaches allow violations of local causality constraint to occur, but are able to detect and recover by rolling back to the point where the violation occurred and reprocessing events in timestamped order.

Geographically distributed simulation models communicate with each other only through message passing, which is generally achieved through middleware. Middleware is connectivity software that contains a set of enabling services that allow multiple processes running on one or more machines to interact across a network. Based on significant standards or products available, middleware can be divided into several categories including Sockets, Remote Procedure Calls (RPC), Remote Method Invocation (RMI), Distributed Object Component Model (DCOM) and Common Object Broker Request Architecture (CORBA) (Tari and Bukhres 2001). During past few years DCOM and CORBA based tools managed to gain a wide acceptability than others.

## 3 PROPOSED APPROACH

### 3.1 Methodology for DMS

The proposed approach for distributed manufacturing simulation includes a simulation methodology and an approach for implementation. In simulation, modeling methodology focuses on the question of how a simulation model should be constructed. A modified version of parallel and distributed simulation methodology presented by Saad et al. (2002), which is shown in Figure 1 will be used as methodology for distributed manufacturing simulation. Most of the activities listed in the proposed methodology are well explained in simulation literature. However, implementation issues such as synchronization, use of middleware need special attention as it is proposed to use commercial simulation software to construct simulation models, and widely available and cost effective technologies to link geographically distributed simulation models.

### 3.2 Synchronization

Synchronization is one of the well-researched areas in parallel and distributed simulation. Optimistic synchronization protocols are implemented by saving the state of the simulation at different time points and allowing the violation of the local causality constraint. If a violation occurs, simulation simply rolls back to a previous time point and resumes again from that particular time point. In conventional distributed simulation this state saving mechanism can be integrated into the simulation engine. However, with commercial simulation software this approach is difficult to implement, as simulation software generally does not allow rolling back to previous time points while it is running. Saving of the simulation state at different time points is also not feasible as users can not access and modify the simulation engine easily. Therefore conservative synchronization protocol is selected for the proposed approach. Defining a value for lookahead is one of the most important aspects of conservative protocols. Unlike with custom built distributed simulations it may not be straightforward to calculate a value for lookahead, as event times are generated by the simulation engine using statistical distributions. However, it is assumed that minimum-processing times for partner organizations can be calculated and these times will be used as lookahead values.

For some distributed simulation systems, particularly distributed manufacturing simulations, a strictly synchronized environment is not required. For these systems an approximate synchronization mechanism is more suitable as approximate approach is simple to implement than strictly synchronized systems. In approximate synchronization ap-proach different models run at different, but approximately close simulation times without using a lookahead.

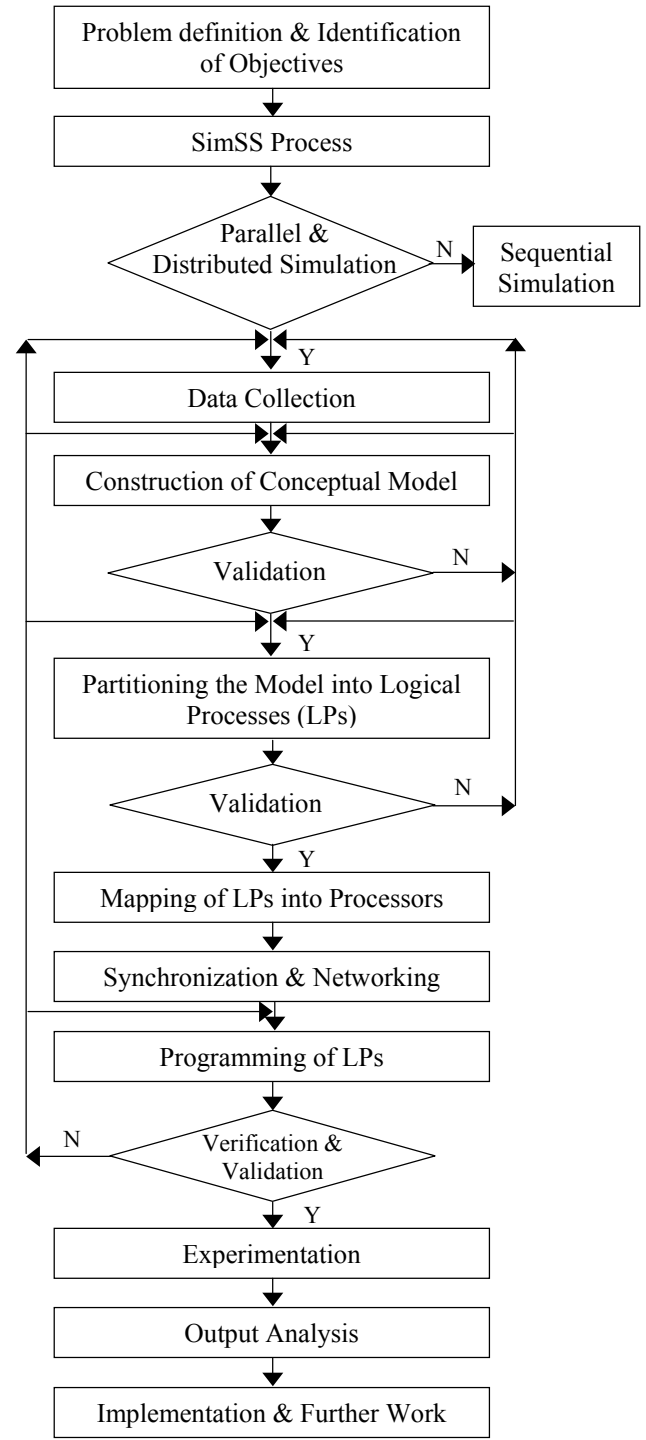

Figure 1: The Proposed Methodology for Distributed Simulation

This is achieved through simulation models comparing simulation times of their own with simulation times of other models. If own simulation time is higher than any other model, it pauses till slower running models reach their simulation time. Since, these simulation models proceed in different time steps, it is impossible to force them to run at exactly same simulation time.

## 3.3 Technology and Software

Analysis of past literature reveals number of attempts to simulated distributed manufacturing systems and supply chains using middleware tools such as HLA, CORBA and GRIDS (see Gan et al. 2000; McLean and Riddick 2000; Taylor et al. 2001; Venkateswaran et al. 2001). For the proposed approach, Microsoft Message Queue (MSMQ), a Message Oriented Middleware was selected to link simulation models. Applications developed with MSMQ can communicate across heterogeneous networks and with computers that may be offline. It provides guaranteed message delivery, efficient routing, security, transactional support, and priority based messaging and could operate in either domain or workgroup environment (Chappell, 1998). In a message queuing system, applications send and receive messages to message queues, which could be located in either local or remote computer. Applications interact with MSMQ via an Application Program Interface (API). MSMQ is integrated into Windows 2000 professional and server editions, as well as Windows XP Professional operating system. It also supports Windows 95, 98 and NT. Windows position as the dominant desktop operating system and increasing acceptability and popularity of windows NT and 2000 server versions as network operating systems led us to select MSMQ, as our motivation was to use widely available and cost effective software and technologies for implementation. MSMQ supports both TCP/IP and IPX network protocols and MSMQ 2.0 can operate with or without MSMQ server. However, if messages are required to route between different domains, a MSMQ server must be configured.

Arena simulation software was used in this study as a commercial package to demonstrate the implementation stage. However, other commercial simulation software such as Automod, Promodel, Witness etc can also be used for this purpose. Visual Basic for Applications (VBA), which offers a programming environment similar to popular Visual Basic, was used to develop the API. Arena and MSMQ support both VBA and C++. Since programming of Arena with VBA is more straightforward, VBA was chosen instead of C++.

## 4 IMPLEMENTATION

To illustrate the implementation of proposed approach, distributed manufacturing model shown in Figure 2 was used.

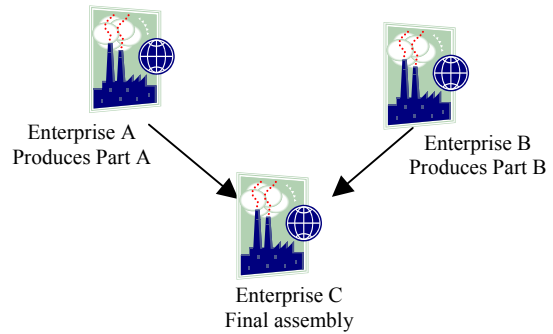

Figure 2: A Model for Distributed Manufacturing Systems

It is assumed that enterprises A and B produce and process Parts A and B respectively and transfer to enterprise C which assembles the product AB. Parts that require rework are send beck to respective producers of parts. In addition to producing and processing Parts A, B and Product AB, companies also produce their own components and products too. It is also assumed that the approximate synchronization approach is used for implementation. For this system, all models are not required to run at approximately same simulation time. However, model C needs to be run at approximately same simulation time of slower running model of either model A or model B.

It was decided to simulate the proposed system in the School of Engineering's main network to emulate the actual implementation environment as far as possible. Since the university uses Novell based system, MSMQ workgroup mode (without a MSMQ server) was selected for implementation. Two MSMQ queues were created for each model. One to receive messages relating to ordering of parts, transferring of parts, returning of defective parts for rework (PartQ), and the other queue to receive messages of synchronization system (TimeQ). Figure 3 shows the composition of a model.

API extracts messages received to queues and passes parameters contained to Arena model, and sends messages to queues of other distributed simulation model containing parameters received from Arena model. When a message arrives to a queue, an event built into API fires automatically to process information included in the message. If message is for incoming batch of parts, API creates and schedules entities in the Arena model.

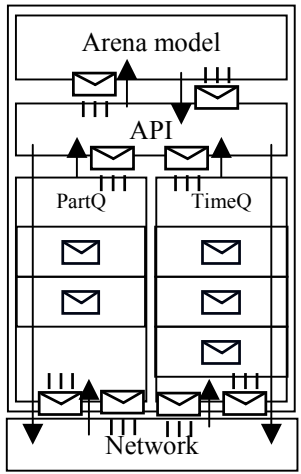

Figure 3: Composition of a Model

Listing of following sample code shows firing an event "qEvent" to create an entity and scheduled to send it after "Delay Time" to "AssignBlock".

```
Option Explicit
Dim qQueue As MSMQQueue
Public WithEvents qEvent As MSMQEvent

Private Sub ModelLogic_RunBeginSimulation()
  Dim qinfo As New MSMQQueueInfo
  qinfo.PathName = ".\private$\PartQ"
  Set qQueue = qinfo.Open(MQ_RECEIVE_ACCESS,_
                          MQ_DENY_NONE)
  Set qEvent = New MSMQEvent
  qQueue.EnableNotification qEvent
End Sub

Sub qEvent_Arrived(ByVal Queue As Object,_
                   ByVal cursor As Long)
  Dim vEntityIndex As Long
  Dim Delay Time As Double
  Dim aQueue As MSMQQueue
  Set aQueue = Queue
  Dim qMsg As MSMQMessage
  Set qMsg = aQueue.Receive(, , , 0)
  vEntityIndex = ThisDocument.Model._
                          SIMAN.EntityCreate
  Call ThisDocument.Model.SIMAN._
       EntitySendToBlockLabel(vEntityIndex,_
                Delay Time, "AssignBlock")
  aQueue.EnableNotification qEvent
End Sub
```

To synchronize the distribute simulation, messages containing simulation times of models A (STa) and B (STb) are sent to model C. Once C receives simulation times from both A and B, it compares its simulation time (STc) with simulation time of slower running model. If simulation time of C is higher than the simulation time of slower model then it pauses till the latter reaches simulation time of C. Figures 4 and 5 show pseudo code of 2 models relating to synchronization. (assuming model A as the slower model).

```
If message label = TIME
  If STc<min(Sta, STb)
    Send STc to slower (A, B) with
       label PAUSE
    Pause model C
  End if
End if
If message label = RESUME
  Resume model C
End if
```

Figure 4: Pseudo Code of Model C

```
If message label = PAUSE
  Schedule an event to send a
    message to C after (STc – STa)
    with label RESUME
End if
```

Figure 5: Pseudo Code of the Slower Model

To measure the effectiveness of the approximate synchronization mechanism, simulation times of all three models were measured against real time clock of the computer with and without the synchronization. Figures 6 and 7 show execution of 3 models with and without approximate synchronization mechanism respectively. It is clear that with the synchronized system, model C runs at approximately same simulation time of slower model from either A or B (Figure 6), and without synchronizing model C runs faster than both models A and B (Figure 7). As simulation time advanced with steps and delays in message passing, it is possible that simulation time of model C may advance to a slightly higher value than other two models before other model force it to pause. This time gap can be varied by changing the inter-arrival time of messages containing simulation times of models A and B.

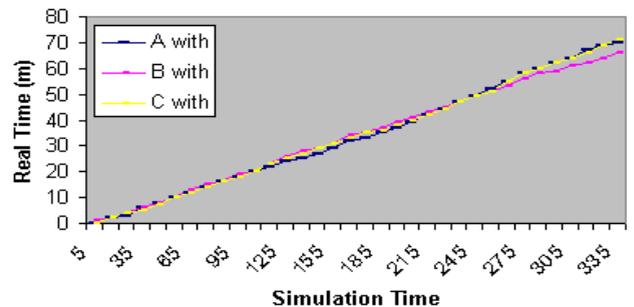

Figure 6: With Approximate Synchronization

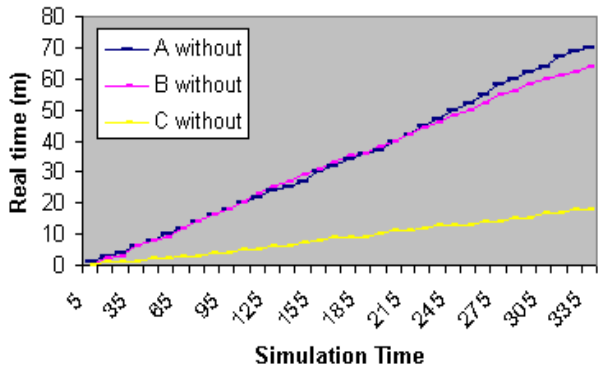

Figure 7: Without Approximate Synchronization

## 5  DISCUSSION

The main aim of this work was to provide a simple and cost-effective solution for simulation of distributed manufacturing enterprises with minimum additional skills needed for the simulation modeler to develop such a system. The proposed system also addresses criticisms leveled towards distributed simulation resulting in a lack of acceptance in general industrial community. Although our main focus was directed towards distributed manufacturing simulation, the systems could also be adapted to simulate non-manufacturing application in distributed simulation environment.

Apart from protection of proprietary information from other companies and additional computational resources provided to the simulation, the approach also encourage reuse of simulation models already developed and used by companies. If an enterprise is already simulating its operations then these models can be linked together by slightly modifying the existing model and adding an API. Cost involved in for additional technologies such as for middleware is minimum as MSMQ is integrated into Windows operating systems and VBA is integrated into simulation packages such as Arena. In addition, programming in VBA does not call for high levels of programming skills too.

One of the most difficult issues to manage the distributed simulation is the synchronization process. If distributed manufacturing simulation does not require to be synchronized strictly, then an approximate synchronization approach can be implemented. This approach works by forcing distributed simulation models to run approximately at same simulation time. However, if strictly synchronized environment is required, the API can easily modified accordingly.

The main benefit of our work was the ability of implementing a distributed simulation with a minimum amount effort and cost. Another advantage is the ability to link simulation models constructed using different commercial simulation software packages, as different companies may be using different simulation software. However, the approach may not work with highly complex systems such as logic circuits, networks, telecommunications systems etc.

## AUTHOR BIOGRAPHIES

**SAMEH M. SAAD** (BSc, MSc, PhD, CEng, MIEE, ILTM) is a Reader in Advanced Manufacturing Systems and Enterprise modeling and management and Postgraduate Course Leader at the Systems and Enterprise Engineering Division, one of the three divisions in the School of Engineering, Sheffield Hallam University, United Kingdom. Dr Saad's research interests revolve around aspects of design and analysis of manufacturing systems, production planning and control, systems integration, reconfigurable manufacturing systems, manufacturing responsiveness, enterprise modeling and management and next generation of manufacturing systems. He has published over 55 articles in various national and international academic journals and conferences. His contact email address is <s.saad@shu.ac.uk>

**TERRENCE PERERA** (BSc, PhD) is Professor and Head of Enterprise and Systems Engineering within the School of Engineering at Sheffield Hallam University. He also leads the Systems Modeling and Integration Research Group. His current research interests include the implementation, integration and practice of virtual modeling tools within all industrial sectors. His email address is <t.d.perera@shu.ac.uk>

**RUWAN WICKRAMARACHCHI** is a PhD student at Sheffield Hallam University, United kingdom. He received his MPil and BSc degrees from University of Cambridge, United Kingdom and University of Kelaniya, Sri Lanka respectively. His main research interest focused on distributed enterprise simulation with emphasis on distributed manufacturing applications. He can be contacted by <w.ruwan@shu.ac.uk>